\documentstyle{article}
\newcommand{\lb}{\label}

\def\buildchar#1#2#3{\null \! \mathop {\vphantom {#1}
\smash #1}\limits ^{#2}_{#3}\!\null }
\def\ut#1{\buildchar{#1}{}{^\sim}\/}

\def\be {\begin{equation}}
\def\ee {\end{equation}}
\def\batwo{\left( \begin{array}{cc}}
\def\bathree{\left( \begin{array}{ccc}}
\def\bafour{\left( \begin{array}{cccc}}
\def\ea{\end{array}\right)}
\def\bea{\begin{eqnarray}}
\def\eea{\end{eqnarray}}

\begin{document}

\begin{titlepage}
\begin{center}
\vfill
{\large\bf  A Connection Approach to Numerical Relativity}
\vfill
{\bf D. C. Salisbury}\footnote{Permanent address: Department of
Physics,
Austin College, Sherman, Texas 75091
\\e-mail: dsalis@austinc.edu}
{\bf and L.C. Shepley}

\vskip 0.4cm
Center for Relativity, The University of Texas at Austin,
\\Austin, TX 78712-1081 USA
\vskip 0.7cm
{\bf Allan Adams}
\vskip 0.4cm
LBJ High School, Austin, Texas USA
\vskip 0.7cm
{\bf Darren Mann and Larry Turvan}
\vskip 0.4cm
Austin College, Sherman, Texas 75091 USA

\vskip 0.7cm
{\bf Brian Turner}
\vskip 0.4cm
Department of Physics, University of Texas at Dallas,
\\Richardson, Texas  USA

\end{center}

\vfill

\begin{center}
{\bf Abstract}
\end{center}
\begin{quote}
We discuss a general formalism for numerically evolving initial
data in general relativity in which  the (complex) Ashtekar connection and
the Newman-Penrose scalars are taken as the
dynamical variables. In the generic case three gauge constraints
and twelve reality conditions must be solved. The analysis is applied
to a Petrov type \{1111\} planar spacetime
where we find a spatially constant volume
element to be an appropriate coordinate gauge choice.
\end{quote}

\vfill

\end{titlepage}

\section{Introduction}

Since their introduction by Abhay Ashtekar in 1986 \cite{Ashtekar86}
\cite{Ashtekar87}, Ashtekar
variables have been the basis of remarkable conceptual and technical
advances toward a quantum theory of gravity  \cite{Ashtekar91}
\cite{Smolin92} \cite{Ashtekar93}. (Br\"{u}gmann maintains
an updated e-mail bibliography on publications related to
Ashtekar variables \cite{Brugmann93}.) Applications have also been found in
classical general relativity. The most fruitful of these have probably
been in the study of Bianchi cosmologies \cite{Kodama88}
\cite{Ashtekar90} \cite{Manojlovic93}. It was suggested soon after
the formalism was discovered that it might lend itself to classical
numerical work, mainly because of the simplicity of the constraints
in terms of the new variables. Meanwhile Capovilla,
Dell, and Jacobson (CDJ) discovered a procedure for
solving algebraically all of the diffeomorphism constraints
\cite{Capovilla89}. This work has opened up new connection-based
approaches to
general relativity. (For further references and an informative schematic
tracing the roots of this current industry to the work of Plebanski
\cite{Plebanski77}, see
the review by Peld\'{a}n \cite{Peldan93}.) Our intention here is to
introduce the CDJ formalism in a general form suitable for undertaking
a finite differencing approach. The program will then be applied in
a testbed which has already served the numerical relativity community
well in developing stable and accurate one-dimensional code:
planar cosmology. Recently, Miller and Smolin have proposed a
discretization scheme which is based on the same (CDJ) action
\cite{Miller93}. The variables and discretization they choose are
likely most suitable for quantum
lattice calculations.

Our plan is first to introduce Ashtekar variables, displaying the
full constraint, evolution, and reality conditions in the 3+1
formalism. In the next section we derive the general dynamical,
constraint, and reality relations using a variation of the (CDJ)
action in which a complex SO(3) connection and a symmetric complex
second rank SO(3) tensor are our variables. These latter objects
turn out to be linear combinations of the five complex Newman-Penrose
scalars, and we present arguments why we might wish to
specify them freely and to monitor their evolution. In Section 3 we
apply the formalism to a spacetime which possesses two commuting
spacelike Killing fields. The simplest such model requires only two
real Newman-Penrose scalars, and is of type \{1111\}. It is not
possible to treat either flat, plane sandwich wave, or colliding
plane waves (prior to collision) in this formalism. Our variable
and coordinate choices are preserved in a spatially constant (but
time dependent) volume gauge.

We conclude with a prognosis of ongoing work in solving numerically
the planar model presented in Section 3 and a discussion of
potential future applications. Appendix A presents a translation
between planar Ashtekar and York variables employed in work by
Anninos, Centrella, and Matzner. In Appendix B we derive the
relation between our scalar variables and the Newman-Penrose scalars.

\section{Ashtekar Variables}

The constraints and evolution equations in the Ashtekar formalism
can be most easily obtained through a variation of the Hilbert
action amended with a term which vanishes automatically by virtue
of the cyclic Bianchi identity \cite{Samuel87} \cite{Jacobson88}.
We first rewrite the Hilbert action in terms
of tetrad fields $E^{\alpha}_{~I}$, using the associated Ricci
rotation coeficients $\Omega_{\mu}^{~I}{}_J$ and curvature
${\cal R}_{\mu \nu~~J}^{~~~I}$
\be
{\cal R}_{\mu \nu~~J}^{~~~I}=2\partial_{{[}\mu}\Omega_{\nu{]~J}}^{~~I}
+2\Omega_{[\mu~M}^{~~I}\Omega_{\nu{]~J}}^{~~M}. \lb{1} \ee
The Ricci scalar ${\cal R}$ is then just
\be
{\cal R}=E^\alpha_{~I}E^\beta_{~J}{\cal R}_{\alpha\beta}^{~~~IJ}
= E^\alpha_{~I}E^\beta_{~J}{\cal R}_{\alpha\beta~K}^{~~~I}
\eta^{JK}. \lb{2}
\ee
The vanishing object which we add is simply $-i$ times the scalar
formed
from the dual $^*{\cal R}_{\mu \nu}^{~~I J}$ of ${\cal R}_{\mu
\nu}^{~~I J}$
\be
^*{\cal R}_{\mu \nu}^{~~I J}:= {1\over2}\epsilon^{I J K L}
{\cal R}_{\mu \nu K L}. \lb{3} \ee

Our conventions are that greek
letters $(\mu, \nu, \rho, ...)$ range over the four spacetime
coordinate indices, while upper case latin letters from the middle of
the
alphabet $(I, J, K, ...)$ range over the four internal SO(1,3) indices.
The SO(1,3) indices are raised and
lowered with the Minkowski metric, which we take to be
$\eta_{I J} = diag(-1,1,1,1)$.
Lower case latin indices from the beginning of the alphabet
$(a, b, c, ...)$ range
over the three spatial indices, while lower case latin indices
from the middle of the alphabet $(i, j, k, ...)$ range over the three internal
SO(3) indices.
Since we shall
frequently deal with concrete indices, we find it useful to
represent contravariant objects with a capital letters, and
corresponding covariant objects with lower case letters. Thus, we
represent the covariant tetrad fields as $e^I_{~\mu}$, with inverse
contravariant field $E^\mu_{~I}$. Where possible we shall distinguish between
four-dimensional objects and three-dimensional objects by
representing the former in either latin script or capital greek letters.

We achieve a suitable 3+1 decomposition by taking $E^{\mu}_{~0}$ to
be
normal to our to our spacelike foliation
\be
E^{\mu}_{~0}=(N^{-1},-N^{-1}N^a), \lb{4} \ee
where $N$ and $N^a$ are the lapse and shift, respectively.
Then
\be
E^{\mu}_{~i} = \delta^{\mu}_a T^a_{~i}, \lb{5} \ee
where $T^a_{~i}$ is the triad field.
The canonical fields in the Ashtekar initial value formulation of
general relativity are the densitized triads $\tilde T^a_{~i}$ and
the connections $A^i_{~a}$.  Let
\be
t := \det t^i_{~a}. \lb{6}
\ee
The densitized triad of weight one, $\tilde T^a_{~i}$, and the densitized
lapse of weight minus one, $\ut N$, are thus
\be
\tilde T^a_{~i} := t T^a_{~i}, \lb{7}
\ee
\be
\ut N := t^{-1} N. \lb{7a}
\ee
We can of course express the metric in terms of $\tilde T^a_{~i}$,
but Rovelli has shown that the densitiized triad itself has
a simple geometrical interpretation:  $\tilde T^a_{~i}$ corresponds
to an SO(3)-Lie-algebra-valued area two-form, and the SO(3) norm of this
two-form
is the area \cite{Rovelli93}.

The Ashtekar connection is constructed from the 3-dimensional
Ricci rotation coefficients $\omega _a^{~i j}$ and the
extrinsic curvature ${\cal K}_{ab}$ in the following manner,
\be
A^i_{~a} :=  \omega _{~a}^i -i K^i_{~a}, \lb{8}
\ee
where
\be
K^i_{~a} := T^{a i}{\cal K}_{ab}, \lb{9}
\ee
and $\omega _{~a}^i$ is the dual of $\omega _a^{~i j}$,
\bea
\omega _{~a}^i &:= &{1\over2} \epsilon^{i j k} \omega _a^{~j k}
\nonumber \\
&= &{1\over2} \epsilon^{i j k}(T^{b[i}t_b{}^{j]}{}_{,a}
-T^{b[i}t_a{}^{j]}{}_{,b}
+ T^{c[i} T^{|b|j]} t_{ak} t_b{}^k{}_{,c}).
\lb{10}
\eea
(Note that since the SO(3) indices are raised and lowered with the
Kronecker delta, we can raise and lower them at our convenience.)

The curvature $F_{a b}^{~~i j}$ associated with the
the Ashtekar connection is
\bea F_{a b}^{~~i j}
&=& 2 \partial_{[ a}A_{b ]}^{~~i j}
+ 2A_{[ a}^{~~i k} A_{b ]}^{~~k j}\nonumber \\
&= & 2 \partial_{[ a}A_{b ]}^{~~i j}
-2 A_{~[ a}^i A_{~b ]}^j,\label{10a} \eea
and the amended Hilbert action $S$ takes the form
\be
S = \int d^4\! x~(i \tilde T^a_{~i} (\dot{ A}^i_{~ a} -
 {\cal D}_a  A^i)
-i N^a \tilde T^b_{~i}F_{a b}^{~ i}
+{1\over2}  \ut N \tilde T^a_{~i}\tilde T^b_{~j}
F_{a b}^{~ i j}). \lb{11} \ee
In (\ref{11}) we have introduced the covariant derivative
operator defined through
\be
{\cal D}_a  A^i := \partial_a A^i + A_a^{~i j} A^j, \lb{12} \ee
where
\be
A^i := \Omega^i_{~0} -i \epsilon^{i j k}\Omega_{0}^{~~0 k}.
\lb{13} \ee

Following the Palatini method, we now vary $\tilde T^a_{~i}$, $A^i_{~a}$,
$\ut N$, $N^a$, and $A^i$ independently in the action (\ref{11}).
The resulting scalar diffeomorphism constraint ${\cal C}$ takes the
form
\be
{\cal C} := \tilde T^a_{~i} \tilde T^b_{~j} F_{a b}^{~~i j} = 0, \lb{14}
\ee
while the vector diffeomorphism constraint ${\cal C}_a$ is
\be
{\cal C}_a := F_{a b}^{~~i} \tilde T^b_{~i} = 0. \lb{15} \ee
We have in addition a gauge constraint ${\cal C}^i$ which reflects or
freedom
to perform local triad rotations
\be
{\cal C}^i := {\cal D}_a \tilde T^a_{~i} = 0. \lb{16} \ee
Our equations of motion are

\be\dot{ A}^i_{~ a} = {\cal D}_a  A^i
+ N^b  F_{b a}^{~ i} + i  \ut N \tilde T^b_{~ j}
 F_{a b}^{~~ i j},\lb{17}\ee
\be\dot{\tilde T}{}^a{}_{~ i}
= \epsilon^{i j k} \tilde T^a_{~ k}  A^j
+ 2 {\cal D}_b N^{[ b} \tilde T^{a ]}_{~ i})
+i\epsilon^{i j k} {\cal D}_b ( \ut N \tilde T^{[ b}_{~ j}
\tilde T^{a ]}_{~ k}).\lb{18}\ee

Finally, we must restrict ourselves to a real metric. A
polynomial form of these reality conditions is achieved by
requiring first that the doubly densitized contravariant metric
$\tilde{\tilde g}{}^{a b} := t^2 g^{ab}$  be real
\be
\Im (\tilde T^a_i \tilde T^b_i ) = 0. \lb{19} \ee
The second reality condition stems ultimately from our
definition (\ref{8}) of the Ashtekar connection. If our
triads were real (as we shall assume below in our planar
example),  we would require that the real part
of the Ashtekar connection be precisely the Ricci
rotation coefficient derived from the triads. It is clear
that this condition cannot involve explicitly a gauge
choice.
(We employ a generalized
notion of gauge which encompasses not only the traditional
internal gauge, but also coordinate choices. So we are
referring to the freedom in $N$ and $N^a$ as well as $A^i$.)
The required conditions can be obtained most efficiently
in polynomial form by requiring that the time derivative of
$\tilde{\tilde g}{}^{a b}$ be real \cite{Ashtekar89a}.
We find that with the choices
$A^i=0$ and $N^a=0$ in the equation of motion (\ref{18}),
\be
\dot{\tilde{\tilde g}}{}^{a b} = 2 i \ut N \epsilon_{i j k} {\cal D}_c
(\tilde T^c_{~j}\tilde T^{(a}_{~~k}) \tilde T^{b)}_{~i}.\lb{19a} \ee
So our second reality condition is
\be
\Re (\epsilon_{i j k} {\cal D}_c
(\tilde T^c_{~j}\tilde T^{(a}_{~~k}) \tilde T^{b)}_{~i} = 0. \lb{20}
\ee

The approach we have described seems at this stage to offer no
practical advantage from a numerical perspective. Indeed, we
may wonder whether any gain at all may be achieved given that
we have increased the number of constraints by three, and we have
twelve non-trivial reality conditions in the generic case.
Nor do the evolution equations
undergo any significant simplifications. Ultimately, of course,
it must be possible to choose gauge conditions such that the
conventional and the Ashtekar numerical approach are exactly
equivalent. One merely constructs the real part of the Ashtekar
connection from the assumed functional form of the real triad fields,
as prescribed in the definition
(\ref{8}). Thus after fulfilling all reality conditions in this manner,
the formalism simply amounts to a change of variables, which may
or may not be useful from the point of view of generating numerical
solutions. This point of view is illustrated in Appendix A in
which we show the equivalence with a numerical construction of
planar cosmologies due to Anninos, Centrella, and Matzner.

Thus a truly new method demands the use of new gauge conditions.
It is
to this strategy we turn next.

\section{A Connection Approach}

The Ashtekar approach was itself stimulated by work in complex
general relativity, and it was from this perspective that a
procedure was discovered by Capovilla, Dell, and Jacobson for
solving the four diffeomorphism constraints (\ref{14}) and
(\ref{15}) \cite{Capovilla89}. They discovered that if $\Psi_{i j}$ is an
arbitrary
traceless, symmetric SO(3) tensor, then the following densitized triad
satisfies the diffeomorphism constraints
\be
\tilde T^a_{~i} = \tilde B^a_{~j} \Psi^{-1}_{~j i}, \lb{21} \ee
where $\tilde B^a_{~j}$ is the double dual of the Ashtekar
curvature
\be
\tilde B^a_{~i} := {1\over4} \epsilon^{a b c}
\epsilon_{i j k} F_{b c}^{~~j k}. \lb{22} \ee
In hindsight this result is not a surprise since the diffeomorphism
constraints are merely algebraic and polynomial in $\tilde T^a_{~i}$ and
$\tilde B^a_{~i}$. Indeed, in substituting our Ansatz (\ref{21})
into the constraints (\ref{14}) and (\ref{15}) we find
\be
{\cal C} = \epsilon_{a b c} \epsilon^{i j k} \tilde T^a_{~i}
\tilde T^b_{~i} \tilde B^c_{~k} = \tilde B \Psi^{-1} \Psi_{i i},
\lb{23} \ee
\be
{\cal C}_a = \epsilon_{a b c} \tilde B^{c i}\tilde T^b_{~i}
= \epsilon_{a b c} \tilde B^{c i} \tilde B^{b j} \Psi_{j i},
\lb{24} \ee
where $\Psi := det (\Psi_{i j})$ and $\tilde B := det (\tilde B^a_{~i})$.
We note in passing that the traceless condition could be
relaxed if we wished to contemplate degenerate metrics, in
which case we would permit $\tilde B = 0$. However, henceforth
we shall assume that $\tilde B \neq 0$.

The components of $\Psi_{i j}$ are simply linear combinations of the
 complex Newman-Penrose scalars $\Psi_i$, where $i=0,...,4$.
In Appendix B we show that
\bea
\Psi_{1 1} &= &{1\over2}(-\Psi_0 + 2 \Psi_2 -\Psi_4), \lb{24a} \\
\Psi_{1 2} &= &{i\over2}(\Psi_0 - \Psi_4), \lb{24b} \\
\Psi_{1 3} &= &\Psi_1 - \Psi_3, \lb{24c} \\
\Psi_{2 2} &= & {1\over2}(\Psi_0 + 2 \Psi_2 +\Psi_4), \lb{24d} \\
\Psi_{2 3} &= & - i \Psi_1 - i \Psi_3, \lb{24e} \\
\Psi_{3 3} &= & - 2 \Psi_2, \lb{24f} \eea
with the inverses,
\bea
\Psi_0 &= &{1\over2}(-\Psi_{1 1} + \Psi_{2 2} - 2 i \Psi_{1 2}),
\lb{24g} \\
\Psi_1 &= &{1\over2}(\Psi_{1 3} +  i \Psi_{2 3}), \lb{24h} \\
\Psi_2 &= &{1\over2}(\Psi_{1 1} + \Psi_{2 2} ), \lb{24i} \\
\Psi_3 &= &{1\over2}(-\Psi_{1 3} + i \Psi_{2 3}), \lb{24j} \\
\Psi_4 &= &{1\over2}(-\Psi_{1 1} + \Psi_{2 2} + 2 i \Psi_{1 2}),
\lb{24k} \eea

These scalars possess a readily measureable physical significance
pointed out by Szekeres \cite{Szekeres65}. $\Psi_0$ and
$\Psi_4$ measure transverse contributions to pure gravitational
waves, while $\Psi_1$ and $\Psi_3$ correspond to longitudinal
contributions. $\Psi_2$ is a Coulombic component which we do not
permit to vanish since we require that $\Psi^{-1}_{~i j}$ exist.
Another quantity of physical interest is the Bel-Robinson
superenergy
tensor
whose positive-definite properties can often be useful
in monitoring gravitational wave disturbances. Its tensorial
expression
in terms of the Newman-Penrose scalars is given in Appendix B.
We view these physical
interpretations as motivation for directly monitoring the
dynamics of the Newman-Penrose scalars.
We must note, however, that in assuming that $\Psi_{ij}$ is invertible,
we are excluding
Petrov types \{31\}, \{4\} and \{-\} from consideration, as we
discuss
 in the Appendix
\cite{Capovilla91} \cite{Penrose86}.

We shall now derive the equations of motion for $\Psi_{i j}$. They
can fact be obtained from the following action \cite{Capovilla91}
\be
S = i \int d^4\! x~  ((\dot{ A}^i_{~ a} -
 {\cal D}_a  A^i)  \tilde B^a_{~j} \Psi^{-1}_{~j i}
+\mu_i \epsilon^{i j k} \Psi_{j k} + \rho \Psi_{i i}). \lb{25}
\ee
(We have added the Lagrange multipliers $\mu_i$ and $\rho$ to
enforce the symmetry
and tracelessness of $\Psi$.)  However, we shall simply substitute
the Ansatz (\ref{21}) for $\tilde T^a_{~i}$ directly into the
equations of motion (\ref{17}) and  (\ref{18}).

The characteristic equation for an arbitrary $3\times3$ matrix
$\Psi_{i j}$ is
\be
\Psi \delta_{i j} = \Psi_{i k} \Psi_{k l} \Psi_{l j}
- \Psi_{k k} \Psi_{i l} \Psi_{l j}
+{1\over2} ((\Psi_{k k})^2 - \Psi_{k l} \Psi_{l k}) \Psi_{i j}. \lb{26} \ee
Thus in our case
\be
\Psi^{-1}_{~i j} = \Psi^{-1}(\Psi_{i k} \Psi_{k j} - {1\over2}
(\Psi_{k k})^2 \delta_{i j}). \lb{27} \ee
Our Ansatz (\ref{21}) for $\tilde T^a_{~i}$
takes the form
\be
\tilde T^a_{~i} = \tilde B^a_{~j} \Psi^{-1}
( \Psi_{j k}  \Psi_{k i} - {1\over2} \delta_{j i}
\Psi_{k l} \Psi_{l k}). \lb{28} \ee
Substituting into the equation of motion (\ref{17}) we find
\bea
\dot{ A}^i_{~ a} &= &{\cal D}_a  A^i
+ \epsilon_{a b c} \tilde B^{b i} N^c   + i  \ut N
\epsilon_{a b c} \epsilon^{i j k} \tilde B^b_{~l} \Psi^{-1}_{~~lj}
\tilde B^c_{~k} \lb{29} \\
&= &{\cal D}_a  A^i
+ \epsilon_{a b c} \tilde B^{b i} N^c   - i  \ut N
\tilde B \Psi^{-1} \tilde b_{a j} \Psi_{j k} \Psi_{k i}. \lb{30} \eea

The equation of motion for $\Psi_{i j}$ requires some more work
and
we will not include all of the details. For this purpose we
require the time derivative of $\tilde B^a_{~i}$
\bea
\dot{\tilde B}{}^a_{~i} &= &{1\over2}\epsilon^{a b c}
\dot{F}_{b c}^{~i}\\
&= &\epsilon^{a b c}{\cal D}_b \dot{A}^i_{~c}. \lb{31} \eea
So according to our Ansatz (\ref{21}) we have
\bea
\dot{\tilde T}{}^a_{~ i} &= &\dot{\tilde B}{}^a_{~j} \Psi^{-1}_{~j i}
-\tilde B^{a j} \Psi^{-1}_{~j m} \dot{\Psi}_{m n} \Psi^{-1}_{~n i}
\lb{32} \\
&=&\epsilon^{j k l} \tilde B^{a l} A^k \Psi^{-1}_{~j i}
+2{\cal D}_b (N^{[b} \tilde T^{a] }_{~~i}) -2 N^{[b}
\tilde B^{a] j} {\cal D}_b \Psi^{-1}_{~~j i} \nonumber \\
&&-i \epsilon^{a b c} {\cal D}_b (\ut N \tilde B \Psi^{-1}
\tilde b_{c k} \Psi_{k i})
+ i \epsilon^{a b c} \ut N \tilde B \Psi^{-1} \tilde b_{c k} \Psi_{k l}
\Psi_{l j} {\cal D}_b \Psi^{-1}_{~~j i} \nonumber \\
&&-\tilde B^{a j} \Psi^{-1}_{~j m} \dot{\Psi}_{m n} \Psi^{-1}_{~n i}.
\lb{33} \eea
In (\ref{33}) we used
\be
F_{b c}^{~~j k} A^k = 2 {\cal D}_{[b} {\cal D}_{c]} A^j \lb{33a} \ee
and the evolution equation (\ref{32}). Comparing the two expressions
for $\dot{\tilde T}{}^a_{~ i}$, (\ref{18}) and (\ref{33}), we
find ultimately that
\be
\dot{\Psi}_{i j} = 2 \epsilon_{k l (i} \Psi_{j) l} A^k
+ N^a {\cal D}_a \Psi_{i j}
- i \ut N {\cal D}_a \Psi_{k j} \epsilon_{i l k} \tilde B^a_{~m}
\Psi^{-1}_{~~m l}. \lb{34} \ee
In deriving (\ref{34}) we made liberal use of the gauge constraint
(\ref{16}). We note here that since
\be
{\cal D}_a \tilde B^a_{~i} \equiv 0, \lb{35} \ee
an equivalent form of the gauge constraint is
\be
{\cal C}_i = \tilde B^a_{~j} {\cal D}_a \Psi^{-1}_{~~j i} = 0. \lb{36} \ee
Notice that $\dot{\Psi}_{i j}$ is manifestly traceless. It is also
symmetric, since
\bea
\epsilon^{i j k}\dot{\Psi}_{i j} &= &-i \ut N
{\cal D}_a \Psi_{k j} \tilde B^a_{~l} \Psi^{-1}_{~~l j}
+ i  \ut N {\cal D}_a \Psi_{j j} \tilde B^a_{~l} \Psi^{-1}_{~~l k}
\nonumber \\
&=&i \ut N \Psi_{k j} \tilde B^a_{~l} {\cal D}_a \Psi^{-1}_{~~l j}
+ 2 i \ut N \epsilon_{m l i} A_{a i} \Psi_{l m}
\tilde B^a_{~j} \Psi^{-1}_{~~j k} = 0. \lb{37} \eea
The first term in the last line vanishes due to the gauge constraint
(\ref{36}), while the second term is zero because of the symmetry
of $\Psi_{i j}$.

\section{Application to Planar Cosmology}

We shall now apply the preceeding formalism to a spacetime which
is presumed to possess two spacelike commuting Killing vectors. In
this case it is possible to introduce a spatial coordinate $x^3=z$
and time $t$ such that the
only nonvanishing densitized triad and connection variables are
$A^1_{~1}$, $A^1_{~2}$, $A^2_{~1}$, $A^2_{~2}$, $A^3_{~3}$, $K^1_{~1}$,
$K^1_{~2}$, $K^2_{~1}$, $K^2_{~2}$, and $K^3_{~3}$ \cite{Husain89}.
We attempt a further simplification motivated by our experience
with conventional variables described in Appendix A. We take the
diagonal components of $A^i_{~a}$ to be pure imaginary, and the
off-diagonal components to be real, so that our non-vanishing
connection components are, in addition to the real
$A^1_{~2}$ and $A^2_{~1}$ (see the definition (\ref{8})),
\be
A^1_{~1} := -i K^1_{~1}, \lb{38} \ee
\be
A^2_{~2} := -i K^2_{~2}, \lb{39} \ee
\be
A^3_{~3} := -i K^3_{~3}, \lb{40} \ee
where the $K$'s are real. The double dual $\tilde B^a_{~i}$
defined in (\ref{22}) is therefore ($~\prime$ is $\partial /\partial z$)
\be \tilde B^a_{~ i} = \bathree - {A_{~2}^1}'
+ K_{~2}^2K_{~3}^3 &
i {K_{~2}^2}'- i A_{~2}^1 K_{~3}^3 & \\
- i {K_{~1}^1}'- i A_{~1}^2 K_{~3}^3 &
{A_{~1}^2}'+ K_{~1}^1 K_{~3}^3 & \\
& & A_{~2}^1A_{~1}^2
+ K_{~1}^1K_{~2}^2 \ea \lb{41}. \ee

Let us take $\Psi_{i j}$ to be diagonal, so that it has the form
\be \Psi = \bathree -a && \\
 &-b& \\
& & a+b \ea \lb{42}, \ee
with $a$ and $b$ real.
Thus we have selected the following nonvanishing
Newman-Penrose scalars
\bea
\Psi_0 &= &{1\over2}(a-b), \lb{42a} \\
\Psi_2 &= & -{1\over2}(a+b), \lb{42b} \\
\Psi_4 &= &{1\over2}(a-b) ~~=~~ \Psi_0. \lb{42c} \eea
This corresponds to a spacetime of Petrov type
\{1111\} \cite{Penrose86}.
We note that we can treat neither plane sandwich
waves nor colliding waves within this formalism
since in both cases there exist open regions which are
of Petrov type \{4\} and/or \{-\}. Our spacetime possesses
equal left and right moving transverse waves, described
by $\Psi_0$ and $\Psi_4$, respectively, and an
everywhere present Coulombic part described by $\Psi_2$. As
mentioned in the introduction, we can now easily monitor
the Bel-Robinson super-energy tensor. In Appendix B, we
show for example, that the super-energy density for observers
in the orthonormal basis $E^\mu_{~I}$ is
\be
T_{0 0 0 0} = {1\over2} (  a^2  + a b + b^2 ). \lb{42d} \ee
Isenberg, Jackson, and Moncrief have investigated the behavior of
this object in a Gowdy $T^3 \times R$ planar spacetime \cite{Isenberg90}.

Then our densitized triads (\ref{21}) are
\be
\tilde T^a_{~i} = \bathree -a^{-1}\tilde B^1_{~1} &
-b^{-1}\tilde B^1_{~2}& \\
-a^{-1}\tilde B^2_{~1} &-b^{-1}\tilde B^2_{~2}& \\
& & (a+b)^{-1}\tilde B^3_{~3}\ea \lb{43}. \ee
We shall
require that $\tilde T^a_{~i}$ be real. Referring to
$\tilde B^a_{~i}$ in (\ref{41}), we are led to our first
realitiy conditions
\be
{K_{~2}^2}'-  A_{~2}^1 K_{~3}^3 = 0, \lb{44} \ee
\be
{K_{~1}^1}'+ A_{~1}^2 K_{~3}^3 = 0. \lb{45} \ee
Thus our densitized triads are diagonal
\be
\tilde T^a_{~i} = \bathree -a^{-1}\tilde B^1_{~1} && \\
&-b^{-1}\tilde B^2_{~2}& \\
& & (a+b)^{-1}\tilde B^3_{~3}\ea \lb{46}, \ee
yielding the metric
\be
g_{a b} = \bathree (\tilde B^1_{~1})^{-1}\tilde B^2_{~2}
\tilde B^3_{~3} b (a+b)
&&\\
&(\tilde B^2_{~2})^{-1}\tilde B^1_{~1} \tilde B^3_{~3} a (a+b)
&\\
&&(\tilde B^3_{~3})^{-1} \tilde B^1_{~1} \tilde B^2_{~2}a b
 \ea .  \lb{46a} \ee

We shall also assume that the only nonvanishing gauge
functions are $A^3:= i {\cal A}^3$, $N^3$, and $\ut N$.
 Substituting
$\tilde T^a_{~i}$ from (\ref{46}) into (\ref{17}) we find
the equations of motion
\be
\dot K^1_{~1} = - {\cal A}^3 A^2_{~1} + \ut N a b^{-1}
(a + b)^{-1} \tilde B^2_{~2} \tilde B^3_{~3}, \lb{47} \ee
\be
\dot K^2_{~2} = {\cal A}^3 A^1_{~2} + \ut N b a^{-1}
(a + b)^{-1} \tilde B^1_{~1} \tilde B^3_{~3}, \lb{48} \ee
\be
\dot K^3_{~3} = {\cal A}^3{}'  + \ut N (a b)^{-1}
(a + b) \tilde B^1_{~1} \tilde B^2_{~2}, \lb{49} \ee
\be
\dot A^1_{~2} = {\cal A}^3 K^2_{~2}- N^3 \tilde B^1_{~1},
\lb{50} \ee
\be
\dot A^2_{~1} = - {\cal A}^3 K^1_{~1}+ N^3 \tilde B^2_{~2}.
\lb{51} \ee

The
remaining equations of motion are (from (\ref{42}) and (\ref{38}))
\be
\dot a = N^3 a'+ \ut N K^2_{~2} \tilde B^2_{~2} b^{-1}
(b + 2 a) + \ut N K^3_{~3} \tilde B^3_{~3} (a-b)
(a-b)^{-1}, \lb{52} \ee
\be
\dot b = N^3 b'+ \ut N K^1_{~1} \tilde B^1_{~1} a^{-1}
(a + 2 b) + \ut N K^3_{~3} \tilde B^3_{~3} (a-b)
(a-b)^{-1}. \lb{53} \ee

Let us next impose the reality condition (\ref{20}). In our
case this is simply the condition that $A^1_{~2}$ and
$A^2_{~1}$ are the only non-vanishing Ricci rotation
coefficients formed from our diagonal densitized triads
(\ref{44}):
\be
A^1_{~2} = \omega^1_{~2} =  {1\over 2 \tilde T^1_{~1}}
\left({\tilde T^3_{~3} \tilde T^1_{~1} \over \tilde T^2_{~2}}\right)'
=-{a \over \tilde B^1_{~1}}\left({a \tilde B^1_{~1}
\tilde B^3_{~3} \over (a+b) a \tilde B^2_{~2}}\right)', \lb{54} \ee

\be
A^2_{~1} = \omega^2_{~1} = -{1\over 2 \tilde T^2_{~2}}
\left({\tilde T^3_{~3} \tilde T^2_{~2} \over \tilde T^1_{~1}}\right)'
={b \over \tilde B^2_{~2}}\left({b \tilde B^2_{~2}
\tilde B^3_{~3} \over (a+b) b \tilde B^1_{~1}}\right)', \lb{55} \ee
It turns out that the gauge constraints ${\cal C}_i=0$ are now
automatically satisfied. This fact is easiest to see referring to
(\ref{16}) and the connection definition (\ref{8}):
\be
{\cal C}_i = \partial_a \tilde T^a_{~i} + \omega_a^{~i j}
\tilde T^a_{~j} - i \epsilon^{i j k} K^k_{~a} \tilde T^a_{~j}
= 0. \lb{56} \ee
The sum of the first two terms vanishes (the densitized
triad is covariantly constant). The second term is
the gauge constraint in a real triad formalism in which the
canonical variables are $\tilde T^a_{~i}$ and $K^i_{~a}$
\cite{Friedman88}. In the present case this constraint vanishes
identically since
both $\tilde T^a_{~i}$ and $K^i_{~a}$ are diagonal.

Next we must insure that our vanishing variable components
remain zero under time evolution. With our choices of gauge
functions, the only non-trivial condition arising from the
evolution of $\Psi_{ij}$ in (\ref{34}) and $A^i_{~a}$ in (\ref{17}) is
\be
-i \dot \Psi_{1 2} = (b-a)({\cal A}^3 - N^3 K^3_{~3})
-\ut N b' \tilde B^3_{~3} (a+b)^{-1}
+\ut N A^1_{~2} \tilde B^2_{~2}b^{-1}(a+2 b) = 0. \lb{57} \ee
Now we must require that the gauge choices (\ref{44}) and
(\ref{45}) are preserved under time evolution. We have
\be
- i \dot{\tilde B}{}^1_{~2} = ({\cal A}^3 - N^3 K^3_{~3})
 \tilde B^1_{~1} - \ut N A^1_{~2} a^{-1} b^{-1} (a+b)
\tilde B^1_{~1} + \left(\ut N b  a^{-1} (a+b)^{-1}
\tilde B^1_{~1} \tilde B^3_{~3}\right)' = 0, \lb{58} \ee
\be
- i \dot{\tilde B}{}^2_{~1} =- ({\cal A}^3 - N^3 K^3_{~3})
 \tilde B^2_{~2} - \ut N A^2_{~1} a^{-1} b^{-1} (a+b)
\tilde B^2_{~2} - \left(\ut N a  b^{-1} (a+b)^{-1}
\tilde B^2_{~2} \tilde B^3_{~3}\right)' = 0, \lb{59} \ee
Two of the preceeding three relations (\ref{57}), (\ref{58}),
and (\ref{59}) fix the gauge functions
${\cal A}^3 - N^3 K^3_{~3}$ and $\ut N$, so the third must be
a new constraint. We present the results without proof as the
calculation is straightforward but tedious. The constraint turns
out to be the requirement that the determinant of the densitized
triads
be spatially constant:
\be
0=\tilde T' = ( \tilde T^1_{~1} \tilde T^2_{~2} \tilde T^3_{~3})'
= \left({ \tilde B^1_{~1}\tilde B^2_{~2}\tilde B^3_{~3}
\over a b (a + b)}\right)'. \lb{60a} \ee
We find that $\ut N$ satisfies the first order
differential equation
\be
\ut N' \ut N^{-1} -A^1_{~2} \tilde B^2_{~2} b^{-1} (a+2b) (a+b)
(b-a)^{-1} + b' \tilde B^3_{~3} (b-a)^{-1} = 0, \lb{61} \ee
so that ${\cal A}^3 - N^3 K^3_{~3}$ may be expressed most
simply as
\be
{\cal A}^3 - N^3 K^3_{~3} = -\ut N' (a+b)^{-1}
\tilde B^3_{~3}. \lb{61a} \ee
The constant volume gauge has been examined by Rovelli in
the context of canonical quantization, although he specializes to
volume elements which are time-independent \cite{Rovelli89}.

Finally, the shift $N^3$ is fixed through the requirement that
the volume element remain spatially constant under time evolution:
\be
0 =\dot{ \tilde T}{}' = 2\left({N^3}' \tilde T + \ut N
\tilde T (K^1_{~1} \tilde T^1_{~1} + K^2_{~2} \tilde T^2_{~2}
+K^3_{~3} \tilde T^3_{~3})\right)', \lb{62} \ee
so that
\be
{N^3}'' = -\left(\ut N ((K^1_{~1} \tilde T^1_{~1} + K^2_{~2} \tilde T^2_{~2}
+K^3_{~3} \tilde T^3_{~3})\right)', \lb{63} \ee
and this equation is readily integrable.

\section{Discussion}

We have developed a general formalism for fixing and evolving
vacuum initial data consisting of a complex SO(3) connection and the
set of Newman-Penrose scalars. Although it is certainly true
that the constraints, reality conditions, and evolution
equations are more complicated than those obtained using conventional
variables, we believe that spacetimes may well exist for which
this complication is compensated by our ability to fix arbitrarily and
then to monitor the time evolution of
the physically significant Newman-Penrose scalars.

We list here the full set of relations obtained in Section 3 for the planar
case. The dynamical variables are $a$, $b$, $K^1_{~1}$, $K^2_{~2}$, and $
K^3_{~3}$, with gauge functions $N^3$, $\ut N$, and ${\cal A}$.
All are real, and they depend only only the coordinates $t$ and $z$. All
constraints are satisfied identically by our dynamical variables, but we have
the following set of reality conditions:
\be
{K'}_{~2}^2-  A_{~2}^1 K_{~3}^3 = 0, \lb{63a} \ee
\be
{K'}_{~1}^1+ A_{~1}^2 K_{~3}^3 = 0, \lb{63b} \ee
\be
A^1_{~2} = \omega^1_{~2} =  {1\over 2 \tilde T^1_{~1}}
\left({\tilde T^3_{~3} \tilde T^1_{~1} \over \tilde T^2_{~2}}\right)'
=-{a \over \tilde B^1_{~1}}\left({a \tilde B^1_{~1}
\tilde B^3_{~3} \over (a+b) a \tilde B^2_{~2}}\right)', \lb{63c} \ee
\be
A^2_{~1} = \omega^2_{~1} = -{1\over 2 \tilde T^2_{~2}}
\left({\tilde T^3_{~3} \tilde T^2_{~2} \over \tilde T^1_{~1}}\right)'
={b \over \tilde B^2_{~2}}\left({b \tilde B^2_{~2}
\tilde B^3_{~3} \over (a+b) b \tilde B^1_{~1}}\right)'. \lb{63d} \ee
The equations of motion are
\be
\dot a = N^3 a'+ \ut N K^2_{~2} \tilde B^2_{~2} b^{-1}
(b + 2 a) + \ut N K^3_{~3} \tilde B^3_{~3} (a-b)
(a-b)^{-1}, \lb{63e} \ee
\be
\dot b = N^3 b'+ \ut N K^1_{~1} \tilde B^1_{~1} a^{-1}
(a + 2 b) + \ut N K^3_{~3} \tilde B^3_{~3} (a-b)
(a-b)^{-1}, \lb{63f} \ee
\be
\dot K^1_{~1} = - {\cal A}^3 A^2_{~1} + \ut N a b^{-1}
(a + b)^{-1} \tilde B^2_{~2} \tilde B^3_{~3}, \lb{63g} \ee
\be
\dot K^2_{~2} = {\cal A}^3 A^1_{~2} + \ut N b a^{-1}
(a + b)^{-1} \tilde B^1_{~1} \tilde B^3_{~3}, \lb{63h} \ee
\be
\dot K^3_{~3} = {{\cal A}'}^3  + \ut N (a b)^{-1}
(a + b) \tilde B^1_{~1} \tilde B^2_{~2}. \lb{63i} \ee
The gauge conditions are
\be
(b-a)({\cal A}^3 - N^3 K^3_{~3})
-\ut N b' \tilde B^3_{~3} (a+b)^{-1}
+\ut N A^1_{~2} \tilde B^2_{~2}b^{-1}(a+2 b) = 0, \lb{63j} \ee
\be
({\cal A}^3 - N^3 K^3_{~3})
 \tilde B^1_{~1} - \ut N A^1_{~2} a^{-1} b^{-1} (a+b)
\tilde B^1_{~1} + \left(\ut N b  a^{-1} (a+b)^{-1}
\tilde B^1_{~1} \tilde B^3_{~3}\right)' = 0, \lb{63k} \ee
\be
({\cal A}^3 - N^3 K^3_{~3})
 \tilde B^2_{~2} + \ut N A^2_{~1} a^{-1} b^{-1} (a+b)
\tilde B^2_{~2} + \left(\ut N a  b^{-1} (a+b)^{-1}
\tilde B^2_{~2} \tilde B^3_{~3}\right)' = 0, \lb{63l} \ee
\be
{N^3}'' = -\left(\ut N ((K^1_{~1} \tilde T^1_{~1} + K^2_{~2} \tilde T^2_{~2}
+K^3_{~3} \tilde T^3_{~3})\right)'. \lb{63m} \ee

We are currently finite-differencing these relations. Our strategy
is to solve the reality conditions for the variables
$K^1_{~1}$, $K^2_{~2}$, and $ K^3_{~3}$ so that the Newman-Penrose scalars
$a$ and $b$ may be specified freely.
We maintain that the complexity
of these relations is in itself not compelling reason to reject
this approach. The issue that counts is whether it is possible to
construct a stable and accurate numerical code, and the jury is
still out.

We anticipate that it might turn out to be useful to combine the
connection approach with the conventional one, possibly in choosing
one technique for evolution and the other for constraints, or
applying both to monitor accuracy. We are also investigating the
possibility of applying the connection approach to the characteristic
initial value problem, where the spinorial formalism has proved
especially useful in analyzing gravitational waves \cite{Penrose86}.

\section{Acknowledgements}

This work was supported by NSF grant number PHY 9211953. D.C.S., A.A., D. M.,
L.T., and B.T. would like to thank the Center for Relativity at The University
of Texas at Austin for its hospitality.

\section*{Appendix A}

We shall give here the change of variables between our variables
$T^a_{~i}$ and $K^i_{~a}$ and the York variables employed by
Anninos, Centrella, and Matzner (ACM) in a series of papers
in which they develop a numerical code for solving Einstein's
equations
in a plane-symmetric spacetime \cite{Anninos89a}
\cite{Anninos89b}
\cite{Anninos91a} \cite{Anninos91b}. The ACM line element is

\be ds^{2}=-{\alpha}^{2}dt^{2}+{\phi^{4}}dx^{2}+
{\phi}^{4}h^{2}dy^{2}+{\phi^{4}}({\beta}dt+dz)^{2}, \lb{b1} \ee
corresponding to the covariant tedrad field
(the components are functions only of $z$ and $t$)
\be
e_{~\mu}^{I}=\bafour \alpha&0&0&0 \\
 0&{\phi^{2}}&0&0 \\ 0&0&{\phi^{2}h}&0 \\
{\beta\phi^{2}}&0&0&{\phi^{2}} \ea. \lb{b2} \ee
The Ashtekar connection is
\be
{A_{~a}^{i}}=\bathree
 i \alpha^{-1}( 2 \beta \phi \phi' -2 \phi \dot\phi )
& \phi^{-1}(2 \phi' h+ \phi h' )&
0 \\ -2\phi^{-1}\phi'&i\alpha^{-1}(2\phi\phi' \beta h +
\phi^2 h' \beta - 2 \phi \dot\phi h - \phi^2\dot h)&0 \\
0&0&i \alpha^{-1}(2 \phi \phi' \beta +  \phi^2  \beta' -
 2 \phi \dot\phi)\ea. \lb{b2a}\ee
Thus
\be
K_{~a}^{i}=\bathree
 -\alpha^{-1}( 2 \beta \phi \phi' -2 \phi \dot\phi )
& 0&
0 \\ 0&-\alpha^{-1}(2\phi\phi' \beta h +
\phi^2 h' \beta - 2 \phi \dot\phi h - \phi^2\dot h)&0 \\
0&0&- \alpha^{-1}(2 \phi \phi' \beta +  \phi^2  \beta' -
 2 \phi \dot\phi)\ea, \lb{b3}\ee
with corresponding equations of motion
\be
\dot \phi = \beta \phi' +{1\over2} \alpha \phi^{-1} K^1_{~1},
\lb{b4} \ee
\be
\dot h = \alpha \phi^{-1} ( K^2_{~2} - K^1_{~1} )
+ h' \beta. \lb{b5} \ee
The remaining equations of motion are
\be
\dot K^1_{~1} =A^2_{~1} {\cal A}^3 +\beta({K^1_{~1}}'+
A^2_{~1}  {K^3_{~3}}')
- \alpha \phi^{-2} h^{-1} ( h {A^2_{~1}}'
+ h K^1_{~1} K^3_{~3}  + A^2_{~1} A^1_{~2}
+K^1_{~1} K^2_{~2}), \lb{b6} \ee
\be
\dot K^2_{~2} =- A^1_{~2} {\cal A}^3 +\beta ({K^2_{~2}}'-
A^1_{~2} K^3_{~3})
- \alpha \phi^{-2}  ( - {A^1_{~2}}'
+ K^2_{~2} K^3_{~3}
+A^2_{~1} A^1_{~2} +
K^1_{~1} K^2_{~2}), \lb{b7} \ee
\be
\dot K^3_{~3} = -{{\cal A}^3}' -\alpha \phi^{-2} h^{-1}
( h {A^2_{~1}}'
+ h K^1_{~1} K^3_{~3} - {A^1_{~2}}' +
 K^2_{~2} K^3_{~3}). \lb{b8} \ee

The non-trivial constraints are the scalar constraint (\ref{15}):
\be
A^2_{~1} A^1_{~2}
+K^1_{~1} K^2_{~2} +
h({A^2_{~1}}' +  K^1_{~1} K^3_{~3})
 - {A^1_{~2}}' +K^2_{~2} K^3_{~3} = 0,\lb{b9} \ee
and the vector constraint ${\cal C}_3$:
\be
h({K^1_{~1}}' + A^2_{~1} K^3_{~3}) +
{K^2_{~2}}' - A^1_{~2} K^3_{~3} = 0 .\lb{b10} \ee

The gauge functions $A^3 := i {\cal A}^3$, $\alpha$, and $\beta$
are partially determined through the conditions that
$\dot{{\tilde T}}{}^1_{~2} = \dot{{\tilde T}}{}^2_{~1}=0$ and
$\dot{{\tilde T}}{}^1_{~1} = \dot{{\tilde T}}{}^3_{~3}$. We find
\be
{\cal A}^3 = -\beta K^3_{~3} + \alpha' \phi^{-2}, \lb{b11} \ee
and
\be
\beta' = \alpha \phi^{-2} (K^3_{~3} - K^1_{~1}). \lb{b12} \ee
As noted in Section 2, there are no further conditions, since
all of our variables are real, and we have constructed the only
nonvanishing $\omega^i_{~a}$'s from the triads:
\be
\omega^1_{~2} = A^1_{~2} = 2 \phi^{-1} \phi' h + h', \lb{b13} \ee
\be
\omega^2_{~1} = A^2_{~1} =  -2 \phi^{-1} \phi' . \lb{b14} \ee

We now give the change of variables to the ACM formalism, where
a York conformal decomposition is employed. ACM define a
traceless extrinsic curvature
\be
A_{a b} := {\cal K}_{a b} - {1\over3} g_{a b} {\cal K}, \lb{b15} \ee
where ${\cal K} := {\cal K}^a_{~a}$. The conformally transformed $ A_{a b}$ is
\be
\hat A_{a b} := \phi^2 A_{a b}. \lb{b16} \ee
The independent variables are taken to be
\be
\hat \eta := \hat A^1_{~1} -  \hat A^2_{~2}, \lb{b17} \ee
in addition to $\hat A^3_{~3}:= \hat A$ and ${\cal K}$. They are obtained
from our variables as follows:
\be
\hat \eta = \phi^4 (  K^1_{~1} - h^{-1} K^2_{~2} ), \lb{b18} \ee
\be
\hat A = {1\over3} \phi^4 ( 2 K^3_{~3} - K^1_{~1}
- h^{-1} K^2_{~2}), \lb{b19} \ee
\be
{\cal K} = \phi^{-2} (  K^1_{~1} + h^{-1} K^2_{~2}  + K^3_{~3}),
 \lb{b20} \ee

Substituting into the constraints (\ref{9}) and (\ref{10})
we find for the scalar constraint
\be
(\phi' h)' = {h \phi\over8}\left( -2 h^{-1} h'' +{2\over3} {\cal K}^2 \phi^4
- {1\over2} ( \hat \eta^2 + 3 \hat A^2 )\phi^{-8}\right),
\lb{b21} \ee
while the vector constraint is
\be
\hat A' - {3\over2} h^{-1} h' \hat A + {1\over2} h^{-1} h' \hat\eta
- {2\over3} \phi^{-2} {\cal K}' = 0. \lb{b22} \ee

As we pointed out in the text, it is by no means obvious that the
system of evolution equations, constraints, and gauge conditions
(\ref{b4}) -  (\ref{b12}) are in any way superior from a numerical
perspective to those obtained through the York procedure. We
have written a fully constrained code in which we solve the constraint
(\ref{b9}) for $K^1_{~1}$ and substitute into the vector constraint
to obtain a first order ordinary differential equation for $K^2_{~2}$.
The metric functions $h$ and $\phi$ are then chosen to represent
incoming colliding waves. Data is evolved using a second-order accurate
leapfrog procedure with downstream updating. This is, however, a
procedure which is also available using conventional variables. Furthermore,
it could be argued that the York procedure possesses greater physical
motivation, since one is free to specify a conformal class of
metrics. Finally, it is much more difficult to implement periodic
boundary conditions with our choice of independent variables. On the other
hand, the connection approach we describe in the text is well motivated
physically, and there is no difficulty in implementing periodic boundary
conditions.

\section*{Appendix B}

We shall show that the components of the symmetric SO(3) tensor
$\Psi_{i j}$ used
in the construction of the densitized triad $\tilde T^a_{~i}$ in
(\ref{21}) are the linear combinations of the Newman-Penrose
scalars given in (\ref{24a} - \ref{24f}). For this purpose it is
most convenient to work with both SU(2) and SL(2,C) spinors.
Although all of the results we shall give here are known, they are
not to be found summarized in one location, and we think our
precedures may also be more accessible to non-experts in spinorial
calculus. In the course of this derivation the relation between
the tetrad and triad 3+1 decomposition will also be made apparent.

 We let $\Sigma^I_{A A'}$ represent $i\over\sqrt{2}$ times the
2x2 identity and Pauli matices ${\bf \Lambda}^i$
\bea
\Sigma^0_{~A A'} &= {i\over\sqrt{2}}{\bf 1} &= {i\over\sqrt{2}}
\batwo
1&0\\0&1 \ea, \\  \lb{a1}
\Sigma^1_{~A A'} &= {i\over\sqrt{2}}{\bf \Lambda}^1
&= {i\over\sqrt{2}}
\batwo
0&1\\1&0 \ea, \\  \lb{a2}
\Sigma^2_{~A A'} &=  {i\over\sqrt{2}}{\bf \Lambda}^2
&= {i\over\sqrt{2}}
\batwo
0&- i\\i&0 \ea, \\  \lb{a3}
\Sigma^3_{~A A'} &= {i\over\sqrt{2}}{\bf \Lambda}^3
&= {i\over\sqrt{2}}
\batwo
1&0\\0&-1\ea. \lb{a4} \eea
The indices $A$ and $A'$ range over 0 and 1. For visualization
purposes we shall conceive of the first index as representing
the row and the second as representing the column. These SL(2,C)
indices may be raised and lowered with the Levi-Civita symbols
\be
\epsilon_{A B} = \epsilon_{A' B'} = \batwo
0&1\\-1&0 \ea. \lb{a5} \ee
Note that we are letting a capital latin letter from the middle of the
alphabet represent a Minkowski index ranging from 0 to 3.

We also introduce SU(2) matrices $\sigma^{i ~A}_{~~~~B} =
\sigma^{i ~~A}_{~B}$ in terms of the Pauli matrices
\bea
\sigma^{1 ~A }_{~~~~B} &=& {i\over\sqrt{2}} \batwo
0&1\\1&0 \ea, \lb{a6} \\
\sigma^{2 ~A }_{~~~~B} &=& {i\over\sqrt{2}} \batwo
0&-i\\i&0 \ea,\lb{a7} \\
\sigma^{3 ~A }_{~~~~B} &=& {i\over\sqrt{2}} \batwo
1&0\\0&-1 \ea,  \lb{a7a} \eea
We have the following useful identities:
\be
\Sigma^I_{~A A'} \Sigma^{J~B A'} \equiv
{1\over2} \eta^{I J} \delta^B_A + ( \delta^{[I}_{~0} \delta^{J]}_k
+{i\over \sqrt{2}} \delta^I_i \delta^J_j \epsilon^{i j k})
\sigma^{k~B}_{~~~~A} , \lb{a8} \ee
\be
\sigma_i^{~A C} \sigma_{j~B C} \equiv {1\over2} \delta_{i j}
\delta^A_B + {1\over \sqrt{2}} \epsilon_{i j k}
\sigma^{k~A}_{~~~~B}. \lb{a9} \ee

With the aid of these matrices and the tetrads given in (\ref{4}) and
(\ref{5}), we can introduce the soldering forms
$\Sigma^\mu_{~A A'}$:
\be
\Sigma^\mu_{~A A'} := E^\mu_{~I} \Sigma^I_{~A A'}. \lb{a10} \ee
As the name suggests, these objects `solder' the tangent space to
spinor space, and we shall now use them to effect our translation
from tensorial to spinor expressions. For this purpose we construct
a complex connection ${\cal A}_\mu^{~~I J}$
\be
{\cal A}_\mu^{~~I J}:= {1\over2}(\Omega_\mu^{~~I J}- i
{}~^*\Omega_\mu^{~~I J}), \lb{a11} \ee
where
\be
^*\Omega_\mu^{~~I J} := {1\over2} \epsilon^{I J K L}
\Omega_{\mu~K L}, \lb{a12} \ee
and $\epsilon^{I J K L}$ is the Levi-Civita symbol with
$\epsilon^{0 1 2 3} = -1$. Since
\be
{}~^*{\cal A}_\mu^{~~I J} = i {\cal A}_\mu^{~~I J}, \lb{a13} \ee
$ {\cal A}_\mu^{~~I J}$ is said to be self-dual. Self-dual objects
have natural spinorial analogues. In particular, we assert without
proof that the spinorial equivalent of the tensorial curvature
${\cal F}_{\mu \nu}^{~~~I J}$ is \cite{Penrose84}
\be
{\cal F}_{\mu \nu}^{~~~I J} =  \Sigma^I_{~A A'}
\Sigma^J_{~B B'} {\cal R}_{\mu \nu}^{~~~A B} \epsilon^{A' B'}
=\Sigma^I_{~A A'} \Sigma^{J~~A'}_{~B} {\cal R}_{\mu \nu}^{~~~A B},
\lb{a14} \ee
where
\be
{\cal F}_{\mu \nu}^{~~~I J} := 2 \partial_{[\mu}{\cal A}_{\nu]}^{~~~I J}
+ 2 {\cal A}_{[\mu}^{~~~I M} {\cal A}_{\nu ]M}{}^J. \lb{a15} \ee
${\cal R}_{\mu \nu}^{~~~A B}$ is the spinorial curvature constructed
from the spinorial connection $\chi_\mu{}^{A B}$ which
leaves the metric $\epsilon_{A B}$ covariantly constant
\be
\chi^{~~A B}_{\mu} = \Sigma^{I~A A'} \Sigma^J_{~B A'}
\Omega_{\mu~I J}. \lb{a16} \ee
Comparing ${\cal A}^i_{~a}$ in (\ref{a11}) with the definition of the
Ashtekar connection $A^i_{~a}$ in (\ref{8}), we discover that
\be
{\cal A}^i_{~a} = {1\over2} A^i_{~a}. \lb{a17} \ee
Thus
\bea
{\cal F}_{a b}^{~~~i j} &= &2 \partial_{[a}{\cal A}_{b]}^{~~~i j}
+ 2 {\cal A}_{[a}^{~~i k} {\cal A}_{b ]k}^{~~~j}
+ 2 {\cal A}_{[a}^{~~i 0} {\cal A}_{b ]0}^{~~~j} \nonumber \\
&= & \partial_{[a}A_{b]}^{~~~i j}
- A_{[a}^i A_{~b ]}^j \lb{a18} \\
&= & {1\over2} F_{a b}^{~~~i j}, \lb{a19} \eea
where in (\ref{a18}) we used the self-duality relation (\ref{a13}).

We are finally prepared to effect the spinorial translation of
(\ref{21}), which we rewrite as
\be
\tilde B^a_{~i} = \Psi_{i j} \tilde T^a_{~j}. \lb{a20} \ee
We have
\bea
\tilde B^a_{~i} &= &{1\over4}\epsilon^{a b c} \epsilon_{i k l}
F_{b c}^{~~k l} \nonumber \\
&= &{1\over\sqrt{2}} \epsilon^{a b c} \sigma^i_{~A B}
{\cal R}_{a b}^{~~A B}, \lb{a21} \eea
where we made use of the identity (\ref{a8}). The
2-form $\Sigma_{a b}^{~~A B}$ will be useful in our
manipulations
\bea
\Sigma_{a b}^{~~A B} &:= &\Sigma_{[a}^{~~A A'}
\Sigma_{b]~~~A'}^{~~B} \nonumber \\
&= &t^i_{~[a} t^j_{~b]} \Sigma_{[i}^{~~A A'}
\Sigma_{j]~~~~A'}^{~~B} \lb{a22} \\
&= &{1\over\sqrt{2}} t^i_{~[a} t^j_{~b]} \epsilon^{i j k}
\sigma^{k~A B} \lb{a23} \\
&= &{1\over\sqrt{2}} \epsilon_{a b c} \tilde T^c_{~i}
\sigma^{i~A B}. \lb{a24} \eea
In (\ref{a22}) we used the covariant tetrad $e^I_{~\mu}$ which
is the inverse of $E^\mu_{~I}$ given in (\ref{4}) and
(\ref{5}):
\be
e^I_{~\mu} = \batwo N&0\\t^i_{~a} N^a & t^i_{~a} \ea, \lb{a25} \ee
while in (\ref{a23}) we used the identity (\ref{a8}) and the
definition (\ref{7}).
We may therefore reexpress  $\tilde T^a_{~i}$ using the identity
(\ref{a9}):
\be
\tilde T^a_{~i} = {1\over\sqrt{2}} \epsilon^{a b c}
\Sigma_{a b}^{~~A B} \sigma_{i~A B}. \lb{a26} \ee
Finally, substituting both (\ref{a21}) and (\ref{a26}) into
(\ref{a20}), we conclude that \cite{Capovilla91}
\bea
{\cal R}_{a b}^{~~C D} &= &\sigma_{i~A B} \sigma_{j~C D}
\Psi_{i j} \Sigma_{a b}^{~~A B} \nonumber \\
&= &\chi_{C D A B} \Sigma_{a b}^{~~A B}, \lb{a27} \eea
where
\be
\chi_{C D A B} := \sigma_{i~A B} \sigma_{j~C D} \Psi_{i j}.
\lb{a28} \ee
Note that because of the symmetry and tracelessness of $\Psi_{i j}$,
$\chi_{A B C D}$ is completely symmetric
\be
\chi_{A B C D} = \chi_{(A B C D)}. \lb{a29} \ee

All that remains is to show that $\chi_{A B C D}$ is simply
the Weyl conformal spinor $\Psi_{A B C D}$. We refer to the
spinorial form of the curvature \cite{Penrose84}
\bea
{\cal R}_{A A' B B' C D} &= &\Sigma^\mu_{~A A'} \Sigma^\nu_{~B B'}
{\cal R}_{\mu \nu}^{~~~C D} \nonumber \\
&= & \Psi_{A B C D} \epsilon_{A' B'}
+ \Phi_{A' B' C D} \epsilon_{A B} \lb{a30} \eea
Thus
\be
{\cal R}_{a b~A B} = \Psi_{A B C D} \Sigma_{a b}^{~~C D}
+ \Sigma_{a~C}^{~~~~A'} \Sigma_b^{~C B'} \Phi_{A' B' A B}. \lb{a31}
\ee
Comparing (\ref{a27}) and (\ref{a31}), we deduce that
\be
\chi_{A B C D} \epsilon_{A' B'} =
\Psi_{A B C D} \epsilon_{A' B'} + \Phi_{A' B' C D} \epsilon_{A B}.
\lb{a32} \ee
It follows that $\Phi_{A' B' C D} = 0$; in other words the Ricci tensor
vanishes, and
\be
\chi_{A B C D} = \Psi_{A B C D} = \sigma_{i~A B} \sigma_{j~C D}
\Psi_{i j}.
\lb{a33} \ee
The definitions of the Newman Penrose scalars are
\bea
\Psi_0 &:= &\Psi_{0 0 0 0}, \lb{a34} \\
\Psi_1 &:= &\Psi_{0 0 0 1}, \lb{a35} \\
\Psi_2 &:= &\Psi_{0 0 1 1}, \lb{a36} \\
\Psi_3 &:= &\Psi_{0 1 1 1}, \lb{a37} \\
\Psi_4 &:= &\Psi_{1 1 1 1}, \lb{a38} \eea
so our claim (\ref{24g}) - (\ref{24k}) has been verified.

We conclude by giving the general tensorial form of the Bel-Robinson
tensor $T_{\mu \nu \rho \sigma}$ \cite{Bel59} \cite{Penrose84}
\be
T_{\mu \nu \rho \sigma} := \Sigma_\mu^{~A A'} \Sigma_\nu^{~B B'}
\Sigma_\rho^{~C C'} \Sigma_\sigma^{~D D'} \Psi_{A B C D}
\overline{\Psi}_{A' B' C' D'}. \lb{a39} \ee
Thus in our tetrad basis the components are
\be
T_{I J K L} = \Sigma_I^{~A A'} \Sigma_J^{~B B'}
\Sigma_K^{~C C'} \Sigma_L^{~D D'} \Psi_{A B C D}
\overline{\Psi}_{A' B' C' D'}. \lb{a40} \ee
We compute, for example, the Bel-Robinson super-energy
density in this orthonormal basis for the planar
cosmology considered in Section 4.
\bea
T_{0 0 0 0} &= &{1\over4}\delta^{A A'}\delta^{B B'}\delta^{C C'}
\delta^{D D'} \Psi_{A B C D}
\overline{\Psi}_{A' B' C' D'} \nonumber \\
&= &{1\over4} (\Psi^2_{~0} + 4 \Psi^2_{~1} + 6 \Psi^2_{~2}
+ 4 \Psi^2_{~3} + \Psi^2_{~4} ) \nonumber \\
&= &{1\over2} (   a^2  +  a b +  b^2 ). \lb{a41} \eea

\end{document}